\documentclass[fleqn,usenatbib]{mnras}

\usepackage[T1]{fontenc}

\DeclareRobustCommand{\VAN}[3]{#2}
\let\VANthebibliography\thebibliography
\def\thebibliography{\DeclareRobustCommand{\VAN}[3]{##3}\VANthebibliography}

\usepackage{graphicx}	
\usepackage{amsmath}	
\usepackage{amssymb}
\usepackage{booktabs}
 
\makeatletter
\newcommand{\citeprocitem}[2]{\hyper@linkstart{cite}{citeproc_bib_item_#1}#2\hyper@linkend}
\makeatother

\newcommand{\G}{\ensuremath{\mathrm{G}}}

\makeatletter
\providecommand*{\diff}{\@ifnextchar^{\DIfF}{\DIfF^{}}}
\def\DIfF^#1{%
    \mathop{\mathrm{\mathstrut d}}\nolimits^{#1}\gobblespace
}
\def\gobblespace{%
    \futurelet\diffarg\opspace
}
\def\opspace{%
    \let\DiffSpace\!%
    \ifx\diffarg(%
        \let\DiffSpace\relax
    \else
        \ifx\diffarg[%
            \let\DiffSpace\relax
        \else
            \ifx\diffarg\{%
                \let\DiffSpace\relax
            \fi\fi\fi\DiffSpace
}

\usepackage{wasysym}

\usepackage{siunitx}
\DeclareSIUnit\parsec{pc}
\DeclareSIUnit\year{yr}
\DeclareSIUnit\week{wk}
\DeclareSIUnit\solarmass{M_{\odot}}
\DeclareSIUnit\arcsecond{as}
\DeclareSIUnit\solarradius{R_{\odot}}
\DeclareSIUnit\solarluminosity{L_{\odot}}

\newcommand{\com}{\ensuremath{\text{COM}}}
\newcommand{\unary}{\ensuremath{\text{u}}}
\newcommand{\bin}{\ensuremath{\text{b}}}
\newcommand{\primary}{\ensuremath{1}}
\newcommand{\secondary}{\ensuremath{2}}
\newcommand{\nunary}{\ensuremath{n_{\text{u}}}}

\newcommand{\nvisual}{\ensuremath{n_{\text{v}}}}
\newcommand{\nvisualprimary}{\ensuremath{n_{\text{v}_{1}}}}
\newcommand{\nvisualsecondary}{\ensuremath{n_{\text{v}_{2}}}}
\newcommand{\nspectroscopic}{\ensuremath{n_{\text{s}}}}
\newcommand{\ntotal}{\ensuremath{n_{\text{total}}}}
\newcommand{\binaryfraction}{\ensuremath{\alpha}}
\newcommand{\spectroscopicfraction}{\ensuremath{\beta}}
\newcommand{\bff}{\ensuremath{\mathrm{B}}}
\newcommand{\ms}[1]{\ensuremath{\sigma^{2}_{#1}}}
\newcommand{\deltams}[1]{\ensuremath{\delta\sigma^{2}_{#1}}}
\newcommand{\visual}{\ensuremath{\text{v}}}
\newcommand{\visualprimary}{\ensuremath{\text{v}_{1}}}
\newcommand{\visualsecondary}{\ensuremath{\text{v}_{2}}}
\newcommand{\spectroscopic}{\ensuremath{\text{s}}}
\newcommand{\spatialresolution}{\ensuremath{\xi}}
\newcommand{\projectedseparation}{\ensuremath{r_{\text{p}}}}
\newcommand{\Projectedseparation}{\ensuremath{R_{\text{p}}}}

\newcommand{\separation}{\ensuremath{r}}

\title{Stellar velocity distributions in binary-rich ultrafaint dwarf galaxies}

\author[A. Gration et al.]{
A. Gration,$^{1}$\thanks{E-mail: a.gration@surrey.ac.uk}
D. D. Hendriks,$^{1}$
P. Das,$^{1}$
D. Heber,$^{1}$
and R. G. Izzard$^{1}$
\\
$^{1}$School of Mathematics and Physics, University of Surrey, Guildford, GU2 7XH \\
}

\date{Accepted XXX. Received YYY; in original form ZZZ}

\pubyear{\the\year{}}

\begin{document}
\label{firstpage}
\pagerange{1--14}
\maketitle

\begin{abstract}
\noindent
Ultrafaint dwarf (UFD) galaxies are dominated by dark matter, the distribution of which may be inferred from the kinematics of that galaxy's stellar population. Star-by-star observations are available for the satellite UFD galaxies of the Milky Way, making them uniquely good laboratories in which to test cosmological predictions at the smallest scales. However, the kinematics of these galaxies are complicated by the presence of binary stars, which alter the stellar velocity distribution. In particular these binary stars increase the galaxy's stellar velocity dispersion, which is related to the total galactic mass by the virial theorem. Without correctly eliminating or accounting for binary stars we may therefore overestimate the masses of UFD galaxies or even confuse globular clusters for UFD galaxies. Here we write down the probability density function for the observed line-of-sight (LOS) velocity of a stellar population containing both visual and spectroscopic binary stars, which we then use to determine the effect of those binary stars on the observed LOS velocity dispersion. For the coldest UFD galaxies the fractional increase in LOS velocity dispersion is of order one and for the coldest globular clusters is of order 100. However, if the stellar initial mass function is bottom light, as it may be for UFD galaxies and globular clusters, then both of these values increase by half a dex.
\end{abstract}


\section{Introduction}

Ultrafaint dwarf (UFD) galaxies have the highest mass-to-light ratios of  all galaxies \citep{simon2019} and are therefore excellent laboratories in which to test cosmological predictions at very small scales. They may be used to constrain the low end of the matter power spectrum \citep{weisz2019} and, since they are less sensitive to feedback than larger dwarf spheroidal (dSph) galaxies, to investigate the primordial density profile \citep{zoutendijk2021} of dark matter halos. UFD galaxies would also provide excellent locations for the indirect detection of dark matter by gamma-ray emmision during the annihilation of weakly interacting massive particles \citep{bengtsson1990,baltz1999}. The density profiles of the dark-matter halos of UFD galaxies may be determined from the kinematics of their stellar populations using dynamical models. The Milky Way's satellite UFD galaxies are close enough for us to make observations of individual stars and are therefore of unique importance.

The construction of a dynamical model of a UFD galaxy is especially simple if we treat its stars as test particles moving in the sum of the potential generated by the dark matter halo and the continuum-field approximation to the potential generated by those stars. It is this approach that we find most commonly in the literature. In particular, we can derive a very simple relationship between mass and velocity dispersion using the virial theorem. Recall that for any gravitationally bound system of stars in equilibrium the virial theorem states that $2\langle{}K\rangle{} + \langle{}U\rangle{} = 0$ where $\langle{K}\rangle $ is the time-averaged total kinetic energy and $\langle{U}\rangle$ is the time-averaged total potential energy. For a population of stars with equal masses moving in a spherical potential it is then the case that 
\begin{align}
\label{eq:virial_theorem}
\ms{V} = \dfrac{\G{}M}{r_{\text{g}}},
\end{align}
where $\ms{V}$ is the stellar velocity dispersion and $r_{\text{g}}$ is the gravitational radius.\footnote{Throughout, we will use the term `velocity dispersion' to mean $\sigma^{2}_{V} := \langle{}V^{2}\rangle{}$.} In practice, this gravitational radius may be approximated by a multiple of the half-light radius, which is an observable quantity, and therefore Equation \ref{eq:virial_theorem} may be used to estimate the masses of galaxies directly. But it also governs the total mass determined by distribution-function models or Jeans models constructed using the same assumptions.

However, if stars exists in binary systems, then the continuum field approximation to the stellar potential fails and the stellar velocity dispersion increases. For the avoidance of doubt let us call the velocity dispersion due to the dark matter halo and continuum field the `intrinsic velocity dispersion'. Binary systems contribute some \emph{additional velocity dispersion}, which we will denote \(\deltams{V}\). When fitting models to observations this additional velocity dispersion appears to contribute an additional mass, \(\delta{}M\). We may then quantify the effect of binary stars on our mass estimates using the \emph{fractional mass increase}, which is 
\begin{align}
\label{eq:fractional_mass_increase}
\dfrac{\delta{}M}{M} = \dfrac{\deltams{V}}{\ms{V}}.
\end{align}

Of course the issue is further complicated by observational practicalities. Most significantly, we are limited to observations of only the line-of-sight (LOS) velocity, \(V_{z}\). This means that we must make some assumption about the symmetry of the velocity distribution. Under the simplest assumption of isotropy the dispersions of all three components of velocity are equal and we have that \(\deltams{V}/\ms{V} = \deltams{V_{z}}/\ms{V_{z}}\). But we must also consider observational lower limits on apparent magnitude and resolution. In failing to resolve the components of a binary star the meaning of its observed LOS velocity is changed (\citeprocitem{8}{El-Badry et al. 2018}). This LOS velocity is determined by fitting models of the star's spectrum to observations. The two components of the binary star together contribute a red- and a blue-shifted line but these lines need not be distinguishable at any one epoch. Moreover the models used often assume that the star in question is part of a unary system and its spectrum therefore has only one line. In these circumstances only one LOS velocity is reported even if the two spectral lines are distinguishable. It need not be the LOS velocity of either component. 

The problem of binary pollution can be largely avoided by using multiepoch observations, which allow us to identify binary stars and then eliminate them from our catalogues \citep{hargreaves1996,minor2010,minor2019}, although the uncertainty in \(\ms{V}\) makes the fractional mass increase uncertain for dynamically cold galaxies in a way it is not for dynamically hotter galaxies. Using multiepoch observations, \cite{martinez1996} found that Segue 1 has an intrinsic LOS velocity dispersion of \(\qty{13.7}{\kilo\meter\squared\per\second\squared}\) while \cite{minor2010} found that Reticulum II has an intrinsic LOS velocity dispersion of \(\qty{7.84}{\kilo\meter\squared\per\second\squared}\). However, for many UFD galaxies we have only one epoch of observations. In these circumstances the additional velocity dispersion can be computed using the statistical distributions of that population's kinematic properties. These distributions are known for the solar neighbourhood and are typically assumed to hold universally.

Using this method \cite{vogt1995} and \cite{minor2010} showed that for moderate binary fractions (of, say, 0.3 or less) robust estimators of the velocity dispersion, which are insensitive to extreme values, can be used to determine the intrinsic velocity dispersion with good accuracy. These robust estimators include the biweight estimator and the 3-sigma velocity clipping algorithm. Typically they produce very similar results. Using 3-sigma velocity clipping \cite{mcconnachie2010} found that the additional velocity dispersion for a population of red-giant branch (RGB) stars could be as great as \(\qty{20}{\kilo\meter\squared\per\second\squared}\). Moreover, they found a greater than 30 per cent chance that the coldest UFD galaxy candidates, Leo IV and Segue 2, are in fact globular clusters with intrinsic velocity dispersions of \(\qty{0.01}{\kilo\meter\squared\per\second\squared}\). However, the binary fraction of UFD galaxies is not well constrained and may be in excess of 0.5 \citep{moe2017} meaning that the use of robust dispersion estimators may be unjustified.

For a complete description of the dynamics of a binary-rich UFD galaxy, however, we would like to have a distribution function model of its phase-space coordinates and observable properties (namely on-sky positions and LOS stellar velocities). Here, we write down the LOS stellar velocity distribution function in its most general form, allowing for distinct subpopulations of unary systems, visual (i.e.\ resolved) binary systems, and spectroscopic (i.e.\ unresolved) binary systems. This allows us compare the kinematics of these stellar subpopulations to their observed kinematics, confounded as they are by resolution limits. It allows us to compute the \emph{expected} additional velocity dispersion and hence the \emph{expected} fractional mass increase. Since UFD galaxies have relatively few bright RGB stars we will in future need to consider fainter stars from the main-sequence turn-off (MSTO). We therefore consider the distribution of LOS velocities for a range of primary-star masses. However, in order to compare our results with those from the literature we also consider a population of binary stars with only RGB primary components. We find that if the binary subpopulation of a UFD galaxy is entirely visual then the additional velocity dispersion can be very large indeed. But if it is partly spectroscopic then the additional velocity dispersion is much reduced. Although we are ultimately interested in the effect of binary systems on the entire distribution of observed LOS stellar velocity we here consider the additional velocity dispersion alone on account of its unique importance. 

Our paper is organized as follows. In Section~\ref{org862a099} we present our distribution-function formalism for the dynamics of UFD galaxies containing a population of binary stars. This section is technical and can be ignored by the reader who is interested only in the results. In Section~\ref{orgf769b1a} we use our formalism to compute the expected additional velocity dispersion and the fractional mass increase resulting from such a population of binary stars. In Section~\ref{org0492588} we discuss these results and compare them to those already in the literature before concluding in Section~\ref{org7dd7a53}.

\section{Methods}
\label{sec:org8f86a11}
\label{org862a099}

The dynamics of a population of stellar systems are completely described by the probability density function (PDF) of its stars' phase coordinates. This is referred to as its `distribution function'. For a population of unary stellar systems we may use this to find the marginal distribution of the observable kinematics, being typically on-sky positions and LOS velocities. However, for resolution-limited observations of a binary-rich stellar population it is not straightforward to perform this marginalization. Here, we therefore consider the PDF of the observed stellar velocities directly and use this to determine the additional LOS stellar velocity dispersion. 

\begin{table}{}
\caption{\label{tab:org2dfcc1a}Physical properties of unary and binary stellar systems. We capitalize these physical properties when treated as random variables. When treated as realizations of random variables we make them lower-case. Velocities are measured with respect to the galactic centre unless they are primed, in which case they are measured with respect to a binary system's COM.
Any quantity $X$ has PDF $f_{X}$ (normalized to one), CDF $F_{X}$, and dispersion $\ms{X}$.}
\centering
\begin{tabular}{p{0.1\linewidth}p{0.8\linewidth}}
\toprule
{Symbol} & {Object}\\[0pt]
\midrule
\(V\) & Velocity of star drawn from galaxy\\[0pt]
\(V_{z}\) & LOS velocity of star drawn from galaxy\\[0pt]
\(V_{\unary, z}\) & LOS velocity of star drawn from population of unary systems\\[0pt]
\(V_{\visual, z}\) & LOS velocity of star drawn from population of visual binary systems\\[0pt]
\(V_{\visual_{1}, z}\) & LOS velocity of primary star drawn from population of visual binary systems\\[0pt]
\(V_{\visual_{2}, z}\) & LOS velocity of secondary star drawn from population of visual binary systems\\[0pt]
\(V_{\spectroscopic, z}\) & LOS velocity of spectroscopic binary star\\[0pt]
\(V_{\primary, z}\) & Velocity of primary star drawn from population of binary systems\\[0pt]
\(V_{\secondary, z}\) & Velocity of secondary star drawn from population of binary systems\\[0pt]
\(L_{\primary}\) & Luminosity of primary star of binary system\\[0pt]
\(L_{\secondary}\) & Luminosity of secondary star of binary system\\[0pt]
\(\Projectedseparation\) & On-sky separation of stars in a binary system\\[0pt]
\bottomrule
\end{tabular}
\end{table}

\begin{table}{}
\caption{\label{tab:org4e3a856}Physical properties of a galaxy.}
\centering
\begin{tabular}{p{0.1\linewidth}p{0.8\linewidth}}
\toprule
{Symbol} & {Object}\\[0pt]
\midrule
\(M\) & Mass\\[0pt]
\(b\) & Characteristic radius\\[0pt]
\(\alpha\) & Binary fraction\\[0pt]
\(\beta\) & Spectroscopic binary fraction\\[0pt]
\(\mathrm{B}\) & Binary fraction factor\\[0pt]
\bottomrule
\end{tabular}
\end{table}

\subsection{The distribution of stellar velocities}

Consider a galaxy consisting of \(n\) stellar systems. Suppose that some number, \(n_{\unary}\), of these systems are unary systems, some number, \(\nvisual\), are visual binary systems, and some number, \(\nspectroscopic\), are spectroscopic binary systems. Then the total number of systems is \(n = \nunary + \nvisual + \nspectroscopic\) and the total number of resolved stars is \(\ntotal = \nunary + 2\nvisual + \nspectroscopic\). We may think of this galaxy as consisting of four stellar populations: one population of unary systems, one population of visual binary-system primary stars, one population of visual binary-system secondary stars, and one population of spectroscopic binary-system stars. Let us denote the number of visual primary stars by \(\nvisualprimary\) and the number of visual secondary stars by \(\nvisualsecondary\) where, of course, \(n_{\visualprimary} = n_{\visualsecondary}\). It will be useful to rewrite these quantities in terms of the galaxy's \emph{binary fraction} (the fraction of stellar systems that are binary), \(\binaryfraction := (\nvisual + \nspectroscopic)/n\), and \emph{spectroscopic binary fraction} (the fraction of binary systems that are spectroscopic), \(\beta := \nspectroscopic/(\nvisual + \nspectroscopic)\). With these definitions in place we have that \(n_{\text{u}} = (1 - \binaryfraction)n\), \(n_{\visualprimary} = n_{\visualsecondary} = \binaryfraction(1 - \spectroscopicfraction)n\), \(n_{\spectroscopic} = \binaryfraction\spectroscopicfraction{}n\) and \(n_{\text{total}} = (1 + \binaryfraction{}(1 - \spectroscopicfraction{}))n\).

The spectroscopic binary fraction is determined by the spatial resolution of our observations. This spatial resolution divides the population of binary systems into two subpopulations: one with projected separations, \(\projectedseparation\), larger than or equal to the spatial resolution, \(\xi\), and one with projected separations smaller than the spatial resolution. LOS velocities for both visual and spectroscopic systems are determined by fitting models to their spectra, typically under the assumption that an unresolved source is a single star \cite[for a discussion, see the paper by][]{el-badry2018}. Thus we observe a single LOS velocity even for a spectroscopic binary system.

If we work in a coordinate system with a $z$-axis coincident with the LOS then a star chosen at random from the population of unary systems will have a LOS velocity, \(v_{\unary, z}\), in the galaxy's centre-of-mass frame. Similarly, stars chosen from the populations of visual primary and visual secondary stars will have LOS velocities \(v_{\visualprimary, z}\) and \(v_{\visualsecondary, z}\). Stars chosen from the population of spectroscopic binary systems will have a LOS velocity \(v_{\spectroscopic, z}\). We can treat these velocities as realizations of random variables, \(V_{\unary, z}\), \(V_{\visualprimary, z}\), \(V_{\visualsecondary, z}\), and \(V_{\spectroscopic, z}\), which have corresponding probability density functions \(f_{V_{\unary, z}}\), \(f_{V_{\visualprimary, z}}\), \(f_{V_{\visualsecondary, z}}\), and \(f_{V_{\spectroscopic, z}}\). The LOS velocity of a star chosen at random from the galaxy \emph{as a whole} can also be treated as a random variable, \(V_{z}\), with PDF \(f_{V_{z}}\). It is equal to \(V_{\unary, z}\), \(V_{\visualprimary, z}\), \(V_{\visualsecondary, z}\), or \(V_{\spectroscopic, z}\) with probabilities
\begin{align}
p_{\unary}
&=\dfrac{(1 - \binaryfraction)}{1 + \binaryfraction{}(1 - \spectroscopicfraction{})},\\
p_{\visualprimary}
&= \dfrac{\binaryfraction(1 - \spectroscopicfraction)}{1 + \binaryfraction{}(1 - \spectroscopicfraction{})},\\
p_{\visualsecondary}
&= \dfrac{\binaryfraction(1 - \spectroscopicfraction)}{1 + \binaryfraction{}(1 - \spectroscopicfraction{})},
\end{align}
and
\begin{align}
p_{\spectroscopic}
&=
\dfrac{\binaryfraction\spectroscopicfraction}{1 + \binaryfraction{}(1 - \spectroscopicfraction{})}.
\end{align}
The distribution of \(V_{z}\) is then a mixture distribution, and the probability density function (PDF) is given by
\begin{align}
\label{eq:pdf_los_velocity}
\begin{split}
f_{V_{z}}(v_{z})
&=
p_{\unary}f_{V_{\unary, z}}(v)
+ p_{\visualprimary}f_{V_{\visualprimary, z}}(v_{z})\\
&\qquad+ p_{\visualsecondary}f_{V_{\visualsecondary, z}}(v_{z})
+ p_{\spectroscopic}f_{V_{\spectroscopic, z}}(v_{z}).
\end{split}
\end{align}

Similary, the components of a binary system chosen at random will have a projected separation that we can treat as the realization of the random variable \(\Projectedseparation\). The spectroscopic binary fraction, \(\spectroscopicfraction\), is the probability that \(\Projectedseparation\), is no greater than \(\spatialresolution\). The spectroscopic binary fraction is then given by the cumulative distribution function (CDF) for \(\Projectedseparation\), namely \(F_{\Projectedseparation}\), such that 
\begin{align}
\label{eq:spectroscopic_fraction}
\spectroscopicfraction
&=
F_{\Projectedseparation}(\spatialresolution).
\end{align}
For the sake of convenience we summarize our notation in Tables~\ref{tab:org2dfcc1a}--\ref{tab:org4e3a856}.

\subsection{The stellar velocity dispersion}
\label{sec:orgacecf55}
\label{org58da4f5}

From the PDF for LOS velocity (Eq.~\ref{eq:pdf_los_velocity}) we immediately find that the LOS velocity dispersion is
\begin{align}
\label{eq:los_stellar_velocity_dispersion}
\begin{split}
\sigma^{2}_{V_{z}}
&=
\dfrac{(1 - \binaryfraction)}{1 + \binaryfraction{}(1 - \spectroscopicfraction{})}\sigma^{2}_{V_{\unary, z}}
+ \dfrac{\binaryfraction(1 - \spectroscopicfraction)}{1 + \binaryfraction{}(1 - \spectroscopicfraction{})}\sigma^{2}_{V_{\visualprimary, z}}\\
&\qquad+ \dfrac{\binaryfraction(1 - \spectroscopicfraction)}{1 + \binaryfraction{}(1 - \spectroscopicfraction{})}\sigma^{2}_{V_{\visualsecondary, z}}
+ \dfrac{\binaryfraction\spectroscopicfraction}{1 + \binaryfraction{}(1 - \spectroscopicfraction{})}\sigma^{2}_{V_{\spectroscopic, z}},
\end{split}
\end{align}
where \(\ms{V_{\unary, z}}\) is the LOS velocity dispersion of the unary systems, \(\ms{V_{\visualprimary, z}}\) is the LOS velocity dispersion of the visual-binary primary stars, \(\ms{V_{\visualsecondary, z}}\) is the LOS velocity dispersion of the visual-primary primary stars, and \(\ms{V_{\spectroscopic, z}}\) is the LOS velocity dispersion of the spectroscopic binary systems. We can rewrite this expression in terms of the inner and outer velocities of the binary systems (i.e.\ the velocities of the binary components in the binary system's rest frame and the velocity of the binary system's COM in the galaxy's rest frame). These, too, are random variables, which we denote \(V'_{\visualprimary, z}\), \(V'_{\visualsecondary, z}\), \(V'_{\spectroscopic, z}\) and \(V_{\visual, z, \com}\), and \(V_{\spectroscopic, z, \com}\) respectively. Let us assume that (1) the outer velocity of any system is independent of its type (be it unary or binary) and (2) the inner velocities of the components of a binary system are independent of the outer velocity. From the first assumption, we have that \(\ms{V_{\unary, z}} = \ms{V_{\visual, z, \com}} = \ms{V_{\spectroscopic, z, \com}}\) and, from the second assumption, we have that
\begin{align}
\ms{V_{\visualprimary, z}}
&= \ms{V_{\visual, z, \com}} + \ms{V'_{\visualprimary, z}},\\
\ms{V_{\visualsecondary, z}}
&= \ms{V_{\visual, z, \com}} + \ms{V'_{\visualsecondary, z}},
\end{align}
and
\begin{align}
\ms{V_{\spectroscopic, z}}
= \ms{V_{\visual, z, \com}} + \ms{V'_{\spectroscopic, z}}.
\end{align}
Let us also introduce the inner LOS velocity of a visual binary system, \(V'_{\visual, z}\), as the LOS velocity of a star drawn at random from a visual binary system. This has PDF
\begin{align}
f_{V'_{\visual, z}} = \dfrac{1}{2}\left(f_{V'_{\visualprimary, z}} + f_{V'_{\visualsecondary, z}}\right)
\end{align}
and dispersion
\begin{align}
\ms{V'_{\visual, z}} = \dfrac{1}{2}\left(\ms{V'_{\visualprimary, z}} + \ms{V'_{\visualsecondary, z}}\right).
\end{align}
We may then rewrite Equation~\ref{eq:los_stellar_velocity_dispersion} as
\begin{align}
\label{eq:mean_square_los_velocity}
\ms{V_{z}} = \sigma^{2}_{V_{\unary, z}} + \deltams{V_{z}},
\end{align}
where
\begin{align}
\label{eq:additional_mean_square_stellar_speed}
\deltams{V_{z}}
=
\dfrac{1}{1 + \binaryfraction{}(1 - \spectroscopicfraction{})}
\left(
2\binaryfraction(1 - \spectroscopicfraction)\sigma^{2}_{V^{\prime}_{\visual, z}}
+ \binaryfraction\spectroscopicfraction
\sigma^{2}_{{V^{\prime}_{\spectroscopic, z}}}
\right).
\end{align}
The first term in Equation~\ref{eq:mean_square_los_velocity} is the galaxy's intrinsic LOS velocity dispersion {in the absence of binary systems} (i.e.\ its LOS velocity dispersion in the absence of binary systems). The second term is the additional LOS velocity dispersion due to the presence of binary systems. Note that, as a result of our assumptions this additional LOS stellar velocity dispersion depends on the properties of the binary-system population and not the properties of the unary-system population.

\subsubsection{The stellar velocity dispersion of resolved and unresolved populations}
\label{sec:orgb22e4d8}

If all binary systems are resolved then \(\spectroscopicfraction = 0\) and the additional LOS stellar velocity dispersion (Eq.~\ref{eq:additional_mean_square_stellar_speed}) becomes
\begin{align}
\label{eq:additional_mean_square_stellar_speed_resolved}
\deltams{V_{z}} = \bff\ms{V^{\prime}_{\visual, z}},
\end{align}
where we have defined the \emph{binary fraction factor} (BFF),
\begin{align}
\label{eq:bff}
\bff := \dfrac{2\binaryfraction}{1 + \binaryfraction}.
\end{align}
This is the fraction of individual stars contained in binary systems, i.e.\ \(\bff = (\nvisualprimary + \nvisualsecondary)/\ntotal\).
Similarly, if all binary systems are unresolved then \(\beta = 1\) and the additional LOS stellar velocity dispersion (Eq.~\ref{eq:additional_mean_square_stellar_speed}) becomes
\begin{align}
\label{eq:additional_mean_square_stellar_speed_unresolved}
\deltams{V_{z}} = \bff\ms{V^{\prime}_{\spectroscopic, z}}.
\end{align}
This last formula was found by \citet[][Eq.\ 9]{minor2010} using different means. 
Note that in each case the additional LOS stellar velocity dispersion consists of two factors:
(1) the BFF, which quantifies the effect of the binary fraction, and
(2) the inner velocity dispersion of the components of the binary systems, which quantifies the effect of the binary stars' orbital properties.

\begin{figure}
\centering
\includegraphics[width=8.4cm]{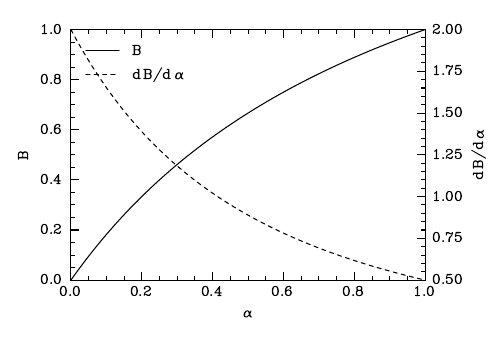}
\caption{
\label{fig:orgd7f9e76}
For a galaxy containing a population of visual binary stars, the additional velocity dispersion is directly proportional to the velocity dispersion of that population (Eq.~\protect\ref{eq:additional_mean_square_stellar_speed_resolved}). The constant of proportionality is the binary fraction factor, \(\mathrm{B}\), which is a function of the binary fraction (i.e.\ the fraction of stellar systems that are binary), \(\alpha\). Its derivative, \(\protect\diff\bff/\protect\diff\binaryfraction\), determines the sensitivity of the additional velocity dispersion to changes in binary fraction (Eq.~\protect\ref{eq:additional_mean_square_stellar_speed_resolved_derivative}).
}
\end{figure}

The effect of the binary fraction is non-linear, i.e.\ the BFF is non-linear in binary fraction, \(\binaryfraction\). Adding more binary systems to the obviously increases the stellar velocity dispersion. But the returns are diminishing. A binary system added to a binary-poor (i.e. low-\(\binaryfraction\)) galaxy has a greater effect on the stellar velocity dispersion than does a binary system added to a binary-rich (i.e. high-\(\binaryfraction\)) galaxy since
\begin{align}
\label{eq:additional_mean_square_stellar_speed_resolved_derivative}
\dfrac{\diff\deltams{V_z}}{\diff\binaryfraction} = \dfrac{\diff\bff}{\diff\binaryfraction}\ms{V^{\prime}_{\visual, z}},
\end{align}
or
\begin{align}
\dfrac{\diff\deltams{V_z}}{\diff\binaryfraction} = \dfrac{\diff\bff}{\diff\binaryfraction}\ms{V^{\prime}_{\spectroscopic, z}},
\end{align}
where
\begin{align}
\label{eq:bff_derivative}
\dfrac{\diff{}\bff}{\diff{}\binaryfraction} = \dfrac{2}{(1 + \binaryfraction)^{2}}.
\end{align}
We plot the BFF and its derivative in Figure~\ref{fig:orgd7f9e76}. Consider, for example, the cases \(\alpha = 0\) and \(\alpha = 1/2\). We see that increasing the binary fraction in the first case has one and a half times the effect on \(\deltams{V}\) than it does in the second case.

\subsection{Computing the stellar velocity dispersion}
\label{sec:org15a3cd7}
\label{org3341d43}

In order to compute the additional LOS velocity dispersion (Eq.~\ref{eq:additional_mean_square_stellar_speed}) we must know the distributions of \(V_{\visualprimary, z}\), \(V_{\visualsecondary, z}\), \(V_{\spectroscopic, z}\), and \(\Projectedseparation\). Recall that the kinematics of a binary system can be described in terms of its component masses (or, equivalently, its primary mass, \(m_{\primary}\), and mass ratio, \(q := m_{\secondary}/m_{\primary}\)) and the orbital elements of the secondary star as observed in the frame comoving with the primary star, namely its semimajor axis, \(a\), eccentricity, \(e\), true anomaly, \(\nu\), argument of the ascending node, \(\Omega\), inclination, \(i\), and argument of pericentre \(\omega\) \citep[see, for example,][]{tremaine2023}.\footnote{Recall that the primary star is defined to be the more luminous of a binary system's two components, having luminosity \(L_{\primary}\), and that the secondary star is defined to be the less luminous, having luminosity \(L_{\secondary}\). For binary systems consisting of ZAMS stars the primary star is always the more massive, so that \(q \in (0, 1]\). However, for old binary systems this is not necessarily the case since the more massive star may have evolved into a compact remnant. In this case the primary star is the less massive of the two components and \(q \in (1, \infty)\). We will call such a binary system `inverted'.} Recall also that $a = (1 + 1/q)a_{\primary} = (1 + q)a_{\secondary}$, where $a_{\primary}$ and $a_{\secondary}$ are the semimajor axes of the primary and secondary stars' orbits in the COM frame.
The components' inner speeds are then
\begin{align}
\label{eq:binary_component_velocity}
{v'}_{\primary}^{2}
= \dfrac{\G{}q^{3}m_{\primary}}{a_{\primary}(1 + q)^{2}}\left(\dfrac{2(1 + e\cos(\nu))}{1 - e^{2}} - 1\right)
\end{align}
and 
\begin{align}
{v'}_{\secondary}^{2} = \dfrac{{v'}_{\primary}^{2}}{q^{2}}
\end{align}
\citep[Eq.\ 1.79][]{tremaine2023}. The components' inner LOS velocities are \begin{align}
\label{eq:los_velocity}
v'_{\primary, z}
&=
\dfrac{h}{l}\sin(i)
\big(
\cos(\nu + \omega) + e\cos(\omega)
\big)
\end{align}
and 
\begin{align}
\label{eq:secondary_los_velocity}
v'_{\secondary, z} = -\dfrac{v'_{\primary, z}}{q}
\end{align}
for specific momentum \(h = (\G{}m_{\secondary}a_{\primary}(1 - e^{2}))^{1/2}/(m_{\primary} + m_{\secondary})\), semilatus rectum \(l = a_{\primary}(1 - e^{2})\), and hence
\begin{align}
\dfrac{h}{l}
=
\left(
\dfrac{\G{}m_{\primary}q^{3}}{a_{\primary}(1 + q)^{2}(1 - e^{2})}
\right)^{1/2}
\end{align}
\cite[Eq.~1.101][]{tremaine2023}. It will be convenient to rewrite this in terms of period, \(p\), as
\begin{align}
\dfrac{h}{l}
=
\dfrac{q}{(1 - e^{2})^{1/2}}
\left(
\dfrac{
2\pi\G{}m_{\primary}
}
{
p(1 + q)^{2}
}
\right)^{1/3}
\end{align}
\cite[Eq.~1.102][]{tremaine2023}.
The inner LOS velocity of a spectroscopic binary can be approximated by the luminosity-weighted average of the LOS velocities of its components
\begin{align}
\label{eq:luminosity_weighted_los_velocity}
v'_{\spectroscopic, z} = \dfrac{L_{\primary}v'_{\primary, z} + L_{\secondary}v'_{\secondary, z}}{L_{\primary} + L_{\secondary}},
\end{align}
where \(L_{\primary}\) and \(L_{\secondary}\) are the luminosities of the primary and secondary star \citep{rastello2020}. The separation of the system's two components is given by the orbit equation, such that \(r = a(1 - e^{2})/(1 + e\cos(\nu))\) (Eq.~1.29, \citeprocitem{29}{Tremaine 2023}). The projected separation is \(\projectedseparation = (x^{2} + y^{2})^{1/2}\) where \(x\) and \(y\) are the on-sky coordinates of one component with respect to the other. By expressing \(x\) and \(y\) in terms of the orbital elements \citep[Eq.~1.70][]{tremaine2023} we find that
\begin{align}
\label{eq:projected_separation}
\projectedseparation
&= \separation\left(\cos^{2}(\nu + \omega) + \cos^{2}(i)\sin^{2}(\nu + \omega)\right)^{1/2}.
\end{align}

A binary system's primary mass, mass ratio, and orbital elements can themselves be treated as random variables, forming the random vector \(\Theta\). The random variables \(V'_{\visualprimary, z}\), \(V'_{\visualsecondary, z}\), \(V'_{\spectroscopic, z}\), and \(\Projectedseparation\) are then functions of \(\Theta\) (by Eqs~\ref{eq:los_velocity}, \ref{eq:luminosity_weighted_los_velocity}, and \ref{eq:projected_separation}). In turn, so are the quantities \(\spectroscopicfraction\), \(\ms{V^{\prime}_{\visual, z}}\), and \(\ms{V^{\prime}_{\spectroscopic, z}}\) (Eq.\ \ref{eq:additional_mean_square_stellar_speed}). 

Even if the joint distribution of $\Theta$ is known in closed form it may not possible to express the distributions of \(V'_{\visualprimary, z}\), \(V'_{\visualsecondary, z}\), \(V'_{\spectroscopic, z}\), and \(\Projectedseparation\) (and hence \(\ms{V^{\prime}_{\visual, z}}\), \(\ms{V^{\prime}_{\spectroscopic, z}}\), and \(\spectroscopicfraction\)) in closed form.
Instead, we must work numerically. To do so, we synthesize the properties of a population of binary systems using Monte Carlo methods and compute the sample velocity dispersions and the sample spectroscopic fraction.

We may, however, say something concrete about the minimum and maximum separation of a binary system's components. Suppose that the distribution of semimajor axes is bounded by \(a_{\min}\) and \(a_{\max}\) and suppose that the eccentricity is bounded by \(e_{\min}\) and \(e_{\max}\). Then the minimum separation occurs at perigee, when it is
\begin{align}
\label{eq:min_separation}
\separation_{\min} = a_{\min}(1 - e_{\max}),
\end{align}
and maximum separation occurs at apogee, when it is
\begin{align}
\label{eq:max_separation}
\separation_{\max} = a_{\max}(1 + e_{\max}).
\end{align}
In turn, the minimum and maximum projected separations are \(r_{\text{p}, \min} = 0\) and \(r_{\text{p}, \max} = {\separation}_{\max}\).

\subsection{The mass function, mass ratio, and orbital elements for zero-age main-sequence systems}
\label{sec:orgd368d2c}
\label{org038997d}

The primary mass, mass ratio, period, and eccentricity have a joint PDF that may be represented as the product of the four PDFs for $Q$ given $(E, P, M_{\primary})$, $E$ given $(P, M_{\primary})$, $P$ given $M_{\primary}$, and $M_{\primary}$:
\begin{align}
\begin{split}
f_{(Q, E, P, M_{\primary})}(q, e, p, m_{\primary}) 
&=
f_{Q|(E, P, M_{\primary})}(q|e, p, m_{\primary})\\
&\qquad{}\times{}f_{E|(P, M_{\primary})}(e|p, m_{\primary})\\
&\qquad{}\times{}f_{P|M_{\primary}}(p|m_{\primary})\\
&\qquad{}\times{}f_{M_{\primary}}(m_{\primary}).
\end{split}
\end{align}
\cite{moe2017} gave closed-form expressions for the first three conditional PDFs in the case of zero-age main-sequence (ZAMS) stars with primary-star masses of $M_{\primary}/M_{\odot} \in [0.8, 40]$.
Their PDFs for $Q$ given $(E, P, M_{\primary})$ and $P$ given $M_{\primary}$ are straightforward to implement. But their PDF for $E$ given $(P, M_{\primary})$ is not. They suggest that for periods equal to or less than some critical value, $p_{\min} = \qty{3.16}{\day}$ (the \emph{circularization period}), all orbits are circular and hence $e = 0$. For periods greater than this they suggest that the eccentricity follows a power-law distribution on the interval $[0, e_{\max}]$ for $e_{\text{max}} = 1 - (p/2)^{-2/3}$, with a power-law index, $\eta$, that is a function of log-period and primary mass, i.e.\ they suggest that
\begin{align}
\label{eq:pdf_eccentricity}
f_{E|P, M_{\primary}}(e|p, m_{\primary})
= A_{E}e^{\eta(\log_{10}(p), m_{\primary})},
\end{align}
where $A_{E}$ is a normalization constant.
However, the formula that they suggest for $\eta$ has the range $(-\infty, 0.9]$ meaning that the integral of the formula given in Equation~\ref{eq:pdf_eccentricity} diverges for $\eta < -1$. It is not, therefore, a valid PDF. We choose to make it valid by restricting the domain of $\eta$ to be that with image $(-1, \infty]$. This is done by excluding some short-period binaries, which we assume to have been circularized.
For a given value of $m_{\primary}$ we may find the new circularization period by inverting the formula given for $\eta$ and evaluating it at $-1$. For $m_{1} \in (0.8, 3]$ we find that $p_{\min} = \qty{8.67}{\day}$, and for $m_{1} \in (7, \infty)$ that $p_{\min} = \qty{4.03}{\day}$. For $m_{1} \in (3, 7]$ the new circularization period lies between these values and must be found numerically.

\begin{figure}[h]
\includegraphics[width=8.4cm]{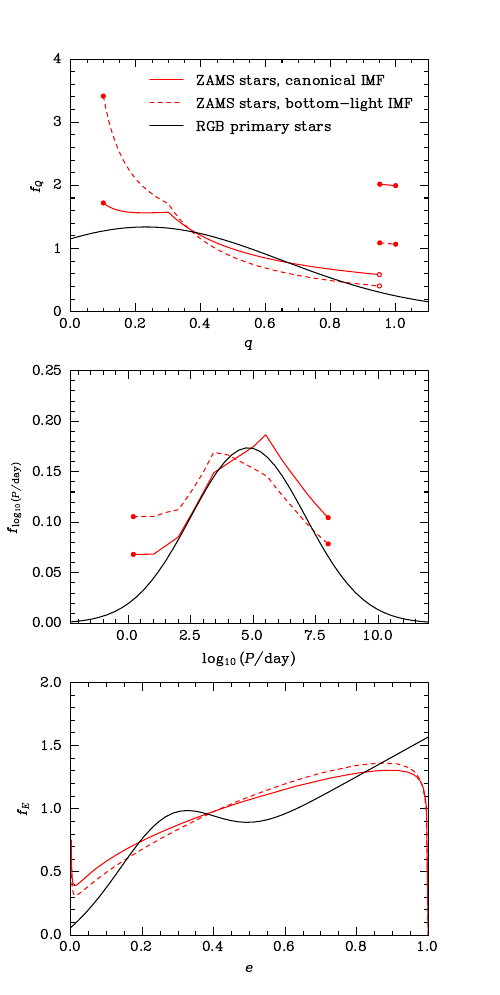}
\caption{
\label{fig:org45f263b}
The velocity dispersion of a population of binary stars is determined by the distributions of primary mass, mass ratio (top), orbital period (middle), and eccentricity (bottom) for a population of binary systems.
We consider three cases:
(1) systems composed of ZAMS stars with primary mass distributed according to the canonical IMF (red, solid),
(2) systems composed of ZAMS stars with primary mass distributed according to a bottom-light IMF (red, dashed), and
(3) systems with RGB primary stars (black, solid).
ZAMS systems with short periods have a greater tendency to be twins, having mass ratio of $q \in (0.95, 1]$, than do long period systems. This explains the discontinuity in the distribution of the mass ratio.}
\end{figure}

To synthesize a population of binary systems, we use the primary-constrained pairing method \citep{kouwenhoven2009}, according to which we assume that the primary-star mass distribution is that of the initial mass function (IMF). 
We take the IMF to be the
Salpeter-like power law so that
\begin{align}
\label{eq:canonical_imf}
f_{M_{\primary}}(m) = A_{M}m_{\primary}^{-2.3}
\end{align}
for 
\begin{align}
A_{M_{\primary}} := \dfrac{1.3}{0.8^{-1.3} - 40^{-1.3}}
\end{align}
on the interval $m_{\primary} \in [0.8, 40]$.
This is the canonical IMF of \cite{kroupa2002} restricted to the mass interval in question.
However, there is some evidence that the IMF for dwarf galaxies is bottom light \citep{geha2013} so we also consider the power law given by
\begin{align}
\label{eq:geha}
f_{M_{\primary}}(m_{\primary})
=
A_{M_{\primary}}m_{\primary}^{-1.3}
\end{align}
for
\begin{align}
A_{M_{\primary}} := \dfrac{0.3}{0.8^{-0.3} - 40^{-0.3}}
\end{align}
again on the interval $m \in [0.8, 40]$.
We plot the PDFs for mass ratio, log-period, and eccentricity (marginalized over all other quantities) in Figure~\ref{fig:org45f263b}. Note that there is an additional population of circularized stars that is not accounted for in the distribution of eccentricity shown here. Below the circularization period the eccentricity of a binary system is always zero and is therefore not a random variable.

Using these distributions we proceed as follows. First, we draw the mass of the primary star using the IMF. Second, we draw the period of the system. Third, we draw the mass of the secondary star. Fourth, we draw the eccentricity of the system. Fifth, we draw the phase of the system using the standard method. This is to say that we draw the mean anomaly of the primary star, \(\mu\), assuming that it is distributed uniformly in the interval \([0, 2\pi)\) and then compute the associated true anomaly using the fact that \(\nu = 2\tan^{-1}((1 + e)^{1/2}\tan(\eta/2)/(1 - e)^{1/2})\) where the eccentric anomaly, \(\eta\), is found by solving Kepler's equation, \(\mu = \eta - e\sin(\eta)\) \citep[Eqs~1.51b and~1.45][]{tremaine2023}. Sixth, we draw the orientation of the system, which we assume to be distributed uniformly on the sphere. Seventh and finally, we determine the luminosities of the two components (required for the luminosity-weighted average LOS velocity for spectroscopic binaries defined by Eq.~\ref{eq:luminosity_weighted_los_velocity}) using the method of \cite{tout1996}.

\subsection{The mass function, mass ratio, and orbital elements for systems with RGB primary stars}
\label{sec:orgf6be682}
\label{org33db866}

For reference we also consider a population of binary systems with RGB primary stars of mass $M_{\primary} = \qty{0.8}{\solarmass}$.
Such a population has been considered in  previous analyses of binary stars' contribution to galaxy dynamics \citep{mcconnachie2010,arroyo-polonio2023}.
In keeping with those analyses we will assume that the mass ratios and orbital elements of these systems are distributed in the same way as they are for present-day systems with solar-type primary stars. These were determined by \cite{duquennoy1991} for late-type class F and early-type class G primary stars with mass  $M_{\primary}/\text{M}_{\odot} \in [0.9, 1.2]$. 
In this narrow range of primary mass the mass ratio and period are effectively independent of all other properties, and eccentricity is effectively conditional only on period.

To synthesize a population of binary systems we proceed by assuming that all systems have a primary-star mass of \(M_{\primary} = \qty{0.8}{\solarmass}\), and then work as we do for ZAMS stars. \cite{duquennoy1991} do not give a closed-form expression for the distribution of eccentricity but we may fit a truncated Gaussian model on the interval \((0, \infty)\) to their data. We find that the least-squares estimates of the parameters \(\mu_{e}\) and \(\sigma^{2}_{e}\) are \(0.27\) and \(0.017\). Since we have no reliable way of computing the luminosity for the secondary stars of inverted binary systems we will truncate this distribution from above at $q_{\max} = 1$. We plot the PDFs for log-period, mass ratio, and eccentricity (again, marginalized over all other parameters) alongside those for \cite{moe2017} in Figure \ref{fig:org45f263b}.

\subsection{The intrinsic velocity dispersion}
\label{sec:orgc549203}
\label{orgd530656}

For our purposes the additional velocity dispersion is only meaningful in comparison with the intrinsic velocity dispersion.
To determine this intrinsic dispersion we may may model the mass distribution of a UFD galaxy's dark matter halo (which dominates its potential) using the Hernquist profile \citep{hernquist1990}.
If we assume the velocity distribution to be isotropic then the intrinsic LOS velocity dispersion is
\begin{align}
\label{eq:hernquist}
\ms{V_{z}}
&= \dfrac{\G{}M}{18b},
\end{align}
where \(M\) is the dynamical mass and \(b\) is a characteristic radius. This formula can be found by direct integration of the phase-space distribution function for a Hernquist profile or by computing its galactic mass and gravitational radius and then using the virial theorem. 
However, our analysis of binary-rich UFD galaxies may also be applied without modification to binary-rich globular clusters, which might be confused for UFD galaxies on dynamical grounds alone \citep{mcconnachie2010}.
It will therefore be worthwhile to consider the intrinsic velocity dispersion of these systems also.
If we model the mass distribution of a globular cluster using a Plummer profile \citep{plummer1911} and again assume the velocity distribution to isotropic then the intrinsic LOS velocity dispersion is
\begin{align}
\label{eq:plummer}
\ms{V_{z}}
&= \dfrac{\pi\G{}M}{32b}.
\end{align}
Note that for both Hernquist and Plummer profiles the intrinsic LOS velocity dispersion is smallest for low-mass diffuse hosts and largest for high-mass concentrated hosts. 

\begin{figure}
\centering
\includegraphics[width=8.4cm]{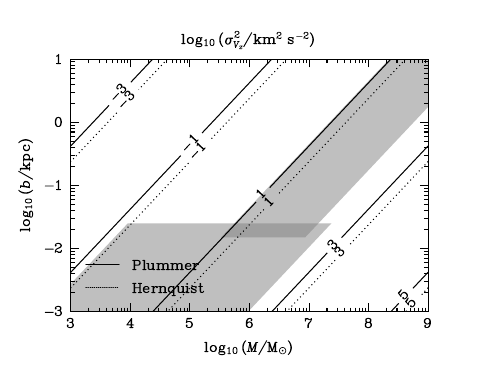}
\caption{\label{fig:orgb074ded}The logarithm of the LOS velocity dispersion, \(\log_{10}(\ms{V_{z}}/\unit{\square\km\per\square\second}\)), for the isotropic Plummer (solid) and isotropic Hernquist (dotted) models as a function of total mass, \(M\), and scale radius, \(b\) (Eqs~\ref{eq:plummer} and~\ref{eq:hernquist}). In both cases, \(\ms{V_{z}} \propto M/b\). For given total mass and scale radius a Plummer-type globular cluster has velocity dispersion larger than does a Hernquist-type galaxy.
The lower (upper) shaded band indicates the region of parameter space consistent with the reported velocity dispersions for the Milky Way's globular clusters (UFD galaxies) assuming a maximum (minimum) half-light radius of \(\qty{20}{\parsec}\).}
\end{figure}

Note, for comparison, that the warmest dSph galaxy, Fornax, has a reported intrinsic LOS velocity dispersion of \(\ms{V_{z}} = \qty{137}{\square\kilo\meter\per\square\second}\) and the coldest UFD galaxy, Leo V, has a reported intrinsic LOS velocity dispersion \(\ms{V_{z}} = \qty{5.29}{\square\kilo\meter\per\square\second}\) \citep{simon2019}. Similarly, the warmest globular cluster, Liller 1, has a reported intrinsic LOS velocity dispersion of \(\ms{V_{z}} = \qty{400.0}{\square\kilo\meter\per\square\second}\) and that the coldest globular cluster, Palomar 14, has a reported intrinsic LOS velocity dispersion \(\ms{V_{z}} = \qty{0.15}{\square\kilo\meter\per\square\second}\) \citep{baumgardt2018} The Milky Way's satellite globular clusters and dwarf galaxies can be distinguished photometrically by their half-light radii. Its globular clusters have half-light radii less than \(\sim \qty{20}{\parsec}\) while its dwarf galaxies have half-light radii greater than \(\sim \qty{20}{\parsec}\) \citep{simon2019}. We plot the velocity dispersions for the Plummer and Hernquist models in Figure~\ref{fig:orgb074ded} along with the regions of parameter space that are consistent with the range of reported velocity dispersions of globular clusters and dwarf galaxies.

\subsection{Implementation}
\label{sec:org4fbcfd0}
\label{org5d10a61}

We implement our numerical method in Python using the library Dyad \citep{gration2025}. Dyad consists of a statistics module that provides a library of distributions for a binary system's physical properties and
a class object that represents a binary system. This class object holds the physical properties of that binary system along with its kinematic properties. For each Monte Carlo realization of a population of binary stars, we consider a sample of size \(10^{6}\).

\section{Results}
\label{sec:org3a37b71}
\label{orgf769b1a}

Having described our formalism for modelling the LOS velocity distribution of clusters and galaxies containing a population of binary stars, we now compute the expected value of the additional LOS velocity dispersion due to this binary-star population. In turn, we consider the expected value of the fractional mass increase for binary-rich UFD galaxies and globular clusters that would arise if all stars were assumed to be in unary systems. We do this for systems consisting of ZAMS stars (Sec.\ \ref{sec:orgd368d2c}) as well as for systems with RGB primaries (Sec.\ \ref{sec:orgf6be682}). We also isolate the contributions of the visual- and spectroscopic-binary subpopulations and determine the effect of a bottom-light IMF on our results.

\subsection{The additional line-of-sight velocity dispersion}
\label{sec:org6583838}

\begin{figure}
\centering
\includegraphics[width=8.4cm]{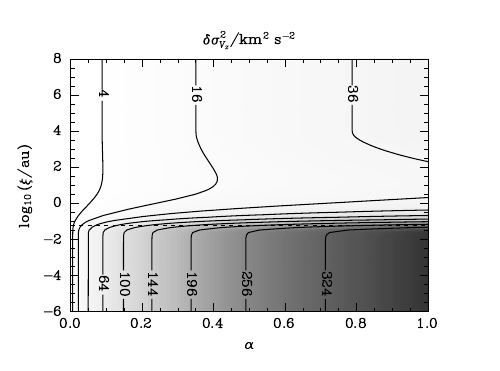}
\caption{\label{fig:org4fe5741}The additional LOS velocity dispersion is a function of binary fraction, \(\alpha\), and spatial resolution, \(\xi\), which determines the sizes of the populations of visual and spectroscopic binary stars (Eq. \ref{eq:mean_square_los_velocity}). Here we show the additional LOS velocity dispersion for binary systems composed of ZAMS stars with masses distributed according to the canonical IMF of \protect\cite{kroupa2002}.}
\end{figure}

\begin{figure}
\centering
\includegraphics[width=8.4cm]{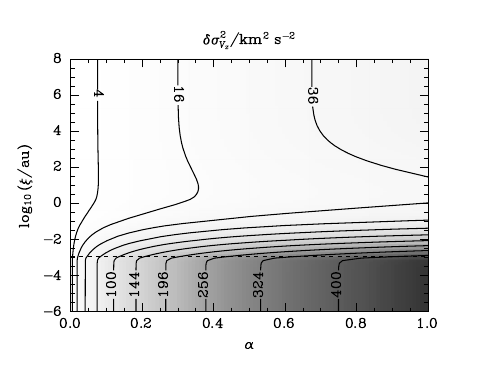}
\caption{\label{fig:org7bdc51c}Same as Figure~\ref{fig:org4fe5741} but for the case of binary systems with RGB primaries.}
\end{figure}

We plot the additional LOS velocity dispersion, \(\deltams{V_{z}}\) (Eq.~\ref{eq:additional_mean_square_stellar_speed}), for our population of ZAMS binary systems in Figure \ref{fig:org4fe5741} using the canonical IMF (Eq.~\ref{eq:canonical_imf}) and for our population of binary systems with RGB primaries in Figure \ref{fig:org7bdc51c}. In each case we see that the additional LOS velocity dispersion is governed by one of two regimes demarcated by a spatial resolution equal to the minimum major axis, \(\xi = 2a_{\min}\). We find that for ZAMS stars \(2a_{\min} = \qty{6.2e-2}{\astronomicalunit}\) and for systems with RGB primaries that \(2a_{\min} = \qty{1.1e-3}{\astronomicalunit}\). 
When \(\xi \ll 2a_{\min}\) nearly all binaries are resolved and we will say that the population is \emph{effectively resolved}. 
When \(\xi \gg 2a_{\min}\) all or nearly all binaries are unresolved and we will say the population is \emph{effectively unresolved}.
For $\xi \ge r_{\max}$ (Eq.~\ref{eq:max_separation}), of course, the population is fully unresolved. We find that for ZAMS stars, $r_{\max} = \qty{1.8e4}{\astronomicalunit}$ and for systems with RGB primaries that $r_{\max} = \qty{8.4e6}{\astronomicalunit}$.

The additional observed LOS velocity dispersion is significantly lower in the effectively unresolved regime than it is in the effectively resolved regime and that within the effectively unresolved regime there is a shallow trough.
When all stars are resolved the additional LOS stellar velocity dispersion is very large indeed. For example, given a binary fraction of \(\alpha = 0.3\), the additional LOS stellar velocity dispersion due to a population of ZAMS binary systems is \(\qty{180}{\square\kilo\meter\per\square\second}\). However observations never fall in this regime. Instead, they always fall in the effectively unresolved regime. Consider angular resolutions of \(\qty{1e-3}{\arcsecond}\) (which is that of \emph{Gaia}) and \(\qty{1e-1}{\arcsecond}\) (which is that of the Subaru Telescope 2.0). The Milky Way's satellites UFD galaxies are at distances of \(20\) to \(\qty{200}{\kilo\parsec}\), corresponding to spatial resolutions of \(20\) to \(\qty{200}{\astronomicalunit}\) or \(2\times{}10^{3}\) to \(\qty{2.e4}{\astronomicalunit}\) respectively. At these resolutions the additional observed LOS velocity dispersion is not only small but nearly constant, especially for small to moderate binary fractions. For example, given \(\alpha = 0.3\) again, the additional LOS velocity dispersion ranges only between \(\qty{12.1}{\square\kilo\meter\per\square\second}\) (in the case of \emph{Gaia}) and \(\qty{13.6}{\square\kilo\meter\per\square\second}\) (in the case of Subaru Telescope 2.0). 

\subsection{The effects of visual and spectrocopic binary stars}
\label{sec:org7180896}

The additional LOS velocity is a function of binary fraction, $\alpha$, spectroscopic binary fraction, $\beta$, and the LOS velocity dispersions for the visual- and spectroscopic-binary subpopulations, $\ms{V^{\prime}_{\visual, z}}$ and $\ms{{V^{\prime}_{\spectroscopic, z}}}$. The last three of these are functions of spatial resolution, $\xi$. In fact, the binary fraction and the spatial resolutions are the two free parameters of our model.
The spatial resolution determines not only the spectroscopic binary fraction but also the distribution of the masses and orbital elements in each of the visual- and spectroscopic-binary subpopulations. These distributions are different from each other and from that of the population as a whole.

\begin{figure}
\centering
\includegraphics[width=8.4cm]{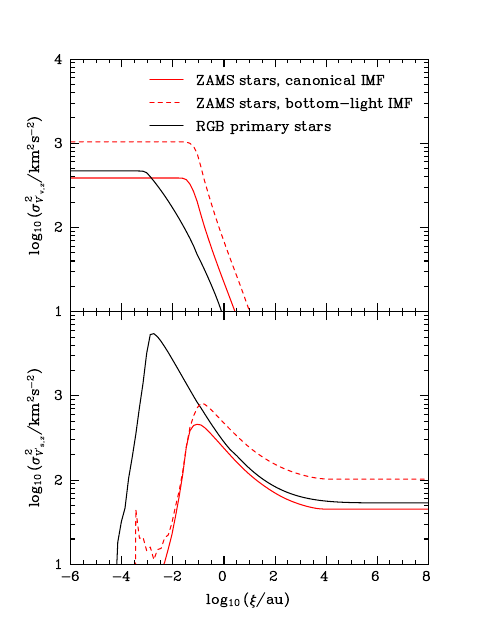}
\caption{\label{fig:org8e8e3eb}
The additional LOS velocity dispersion is a linear combination of the LOS velocity dispersions of the populations of visual (top panel) and spectroscopic (bottom panel) binary stars, $\ms{V_{\visual, z}}$ and $\ms{V_{\spectroscopic, z}}$ (Eq.~\ref{eq:additional_mean_square_stellar_speed}). Here we show these velocity dispersions for our three cases:
(1) systems composed of ZAMS stars with primary mass distributed according to the canonical IMF (red, solid),
(2) systems composed of ZAMS stars with primary mass distributed according to a bottom-light IMF (red, dashed), and
(3) systems with RGB primary stars (black, solid).}
\end{figure}

We plot $\ms{V^{\prime}_{\visual, z}}$ and $\ms{{V^{\prime}_{\spectroscopic, z}}}$ as functions of $\xi$ in Figure~\ref{fig:org8e8e3eb}. Note that $\ms{V^{\prime}_{\visual, z}}$ is large and constant in the effectively resolved regime (where $\xi \ll 2a_{\min}$) and is decreasing in the effectively resolved regime (where $\xi \gg 2a_{\min}$).
However, $\ms{V^{\prime}_{\spectroscopic, z}}$ is peaked at the boundary between the two regimes. It increases in the effectively resolved regime and is small and constant in the effectively unresolved regime.

Consider a fully resolved population for which $\xi$ vanishes.
As $\xi$ increases the visual-binary subpopulation consists of more and more systems with long periods, high eccentricities, and low mass ratios, resulting in smaller and smaller and LOS velocity dispersion.
Contrariwise, consider a fully unresolved population for which $\xi = r_{\max}$. 
The LOS velocities of the two components in each system completely or partially cancel each other (Eq.~\ref{eq:luminosity_weighted_los_velocity}) so that the LOS velocity dispersion is significanly smaller than it is for a fully resolved population.
As $\xi$ decreases the spectroscopic-binary subpopulation consists of more and more systems with short periods, low eccentricities, and high mass ratios.
Alongside these changes, more and more of the spectroscopic systems are viewed edge-on or nearly edge-on.
This causes the LOS velocity dispersion to increase until $\xi = 2a_{\min}$, below which it then decreases.

\begin{figure}
\centering
\includegraphics[width=8.4cm]{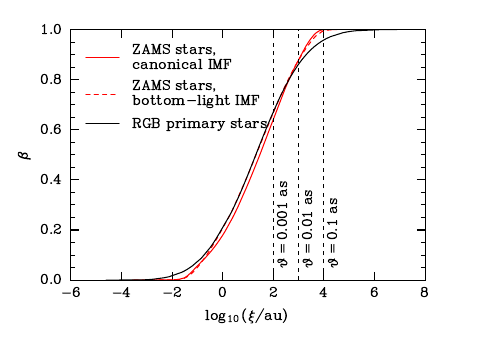}
\caption{\label{fig:org1358c78}The sizes of the populations of visual and spectroscopic binary stars are determined by the spatial resolution. The spectroscopic binary fraction (i.e.\ the fraction of binary stars that are spectroscopic) is given by the CDF for projected separation, \(\projectedseparation\). Here we show this CDF for our three cases. The vertical lines indicate the spatial separation threshold for a galaxy at \(\qty{100}{\kilo\parsec}\) assuming angular resolution of \(\theta = \qty{0.001}{\arcsecond}, \qty{0.01}{\arcsecond}, \qty{0.1}{\arcsecond}\).}
\end{figure}

We plot $\beta$ as a function of $\xi$ in Figure~\ref{fig:org1358c78}, where we have marked spatial resolutions of \(\xi = \qty{1e2}{\astronomicalunit}\), \(\xi = \qty{1e3}{\astronomicalunit}\), and \(\xi = \qty{1e4}{\astronomicalunit}\), which correspond to angular resolutions of \(\theta = \qty{1e-1}{\arcsecond}\), \(\theta = \qty{1e-2}{\arcsecond}\), and \(\theta = \qty{1e-3}{\arcsecond}\) for a system at a distance of \(\qty{100}{\kilo\parsec}\).
For these spatial resolutions the spectroscopic binary fractions are 1, 0.88, and 0.64 respectively in the case of ZAMS stars with canonical IMF and 0.96, 0.86, and 0.67 respectively in the case of systems with RGB primaries.

\subsection{The fractional mass increase}
\label{sec:org982f264}

\begin{figure}
\centering
\includegraphics[width=8.4cm]{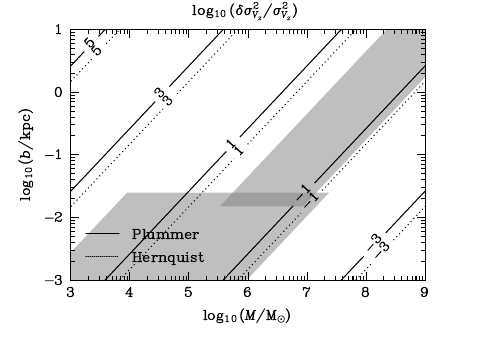}
\caption{\label{fig:org1ea136d}The logarithm of the fractional increase in velocity dispersion, \(\log_{10}(\deltams{V_{z}}/\ms{V_{z}})\), for a Plummer-type globular cluster or Hernquist-type UFD galaxy with binary fraction \(\binaryfraction = 0.3\) observed with angular resolution \(\theta = \qty{1.e-2}{\as}\) at a distance of \(\qty{100}{\kilo\parsec}\) and populated with ZAMS stars assuming the initial mass function of Kroupa. The lower (upper) shaded band indicates the region of parameter space consistent with the reported velocity dispersions for the Milky Way's globular clusters (UFD galaxies) assuming a maximum (minimum) half-light radius of \(\qty{20}{\parsec}\).}
\end{figure}

We now compute the fractional increase in LOS velocity dispersion, \(\deltams{V_{z}}/\ms{V_{z}}\), that would result from assuming that the stellar population is made up only of unary systems. This is equal to the fraction mass increase, $\delta{}M/M$ (Eq.~\ref{eq:fractional_mass_increase}), if we assume the velocity distribution of the cluster or galaxy to be isotropic. Let us assume a binary fraction \(\alpha = 0.3\) and spatial resolution \(\projectedseparation = \qty{1e3}{\astronomicalunit}\), corresponding to angular resolution of \(\theta = \qty{1e-2}{\arcsecond}\) at a distance of \(\qty{100}{\kilo\parsec}\). For ZAMS stars the additional LOS velocity dispersion is \(\deltams{V_{z}} = \qty{13.1}{\square\kilo\meter\per\square\second}\) while for systems with RGB primaries the additional velocity dispersion is \(\deltams{V_{z}} = \qty{15}{\square\kilo\meter\per\square\second}\). We plot the resulting fractional mass increase as a function of host mass, \(M\), and host scale radius, \(b\), for the Plummer sphere and Hernquist model (Eqs~\ref{eq:plummer} and~\ref{eq:hernquist}) using our population of ZAMS binary systems in Figure~\ref{fig:org1ea136d}.

By considering the regions of parameter space consistent with reported velocity dispersions (Sec.~\ref{sec:orgc549203}) we find that UFD galaxies exhibit fractional increases in LOS velocity dispersion of order 0.1 to 1 and that globular glusters exhibit fractional increases in LOS velocity dispersion of order 0.1 to 100. The same is true in the case of a binary population consisting of systems with RGB primaries.

\subsection{The impact of a bottom-light initial mass function}
\label{sec:org1274f71}
\label{orgd0e7ed2}

\begin{figure}
\centering
\includegraphics[width=8.4cm]{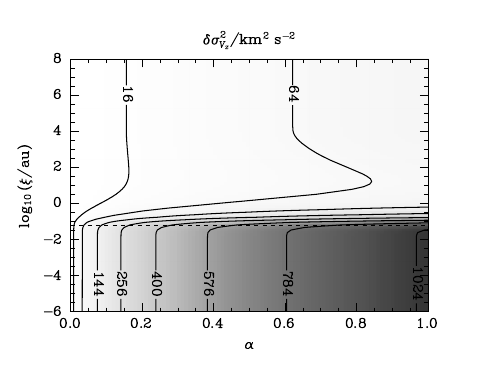}
\caption{\label{fig:orga7a7ff1} Same as Figure~\ref{fig:org4fe5741} but for a bottom-light IMF.}
\end{figure}

\begin{figure}
\centering
\includegraphics[width=8.4cm]{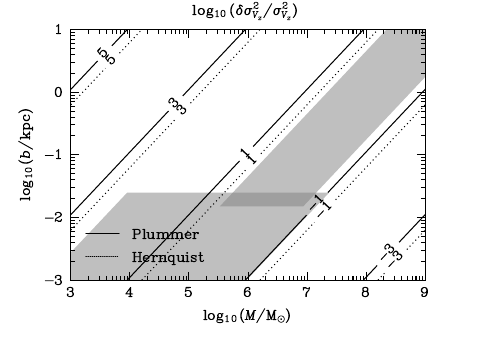}
\caption{\label{fig:org016fd89} Same as Figure~\ref{fig:org1ea136d} but for a bottom-light IMF. The additional velocity dispersion is everywhere increased by approximately 0.5 dex.}
\end{figure}

We plot the additional LOS velocity dispersion, \(\deltams{V_{z}}\), for our population of ZAMS binary systems using a bottom-light IMF (Eq.~\ref{eq:geha}) in Figure~\ref{fig:orga7a7ff1}. In these circumstances the additional LOS velocity dispersion is everywhere significantly greater than it is when using the canonical IMF. For a binary fraction \(\alpha = 0.3\) and spatial resolution \(\projectedseparation = \qty{1e3}{\astronomicalunit}\) we find that the additional LOS velocity dispersion is \(\deltams{V_{z}} = \qty{36.0}{\square\kilo\meter\per\square\second}\).
We plot the resulting fractional increase in LOS velocity dispersion as a function of host mass, \(M\), and characteristic radius, \(b\), for the Plummer sphere and Hernquist model in Figure~\ref{fig:org016fd89}. These results differ from those in the case of the canonical IMF by half a dex.

\section{Discussion}
\label{sec:orgfae4374}
\label{org0492588}

In interpreting our results it is worth considering three factors:
the distributions of primary mass and orbital elements,
the definition of the inner LOS velocity of a spectroscopic binary, and
the assumptions we have made in deriving our expression for the additional LOS velocity dispersion. The additional LOS velocity dispersion is sensitive to each of these.

\subsection{The distributions of primary mass and orbital elements}
\label{sec:orgb6b717a}
\label{org1b49edb}

\begin{figure}
\centering
\includegraphics[width=8.4cm]{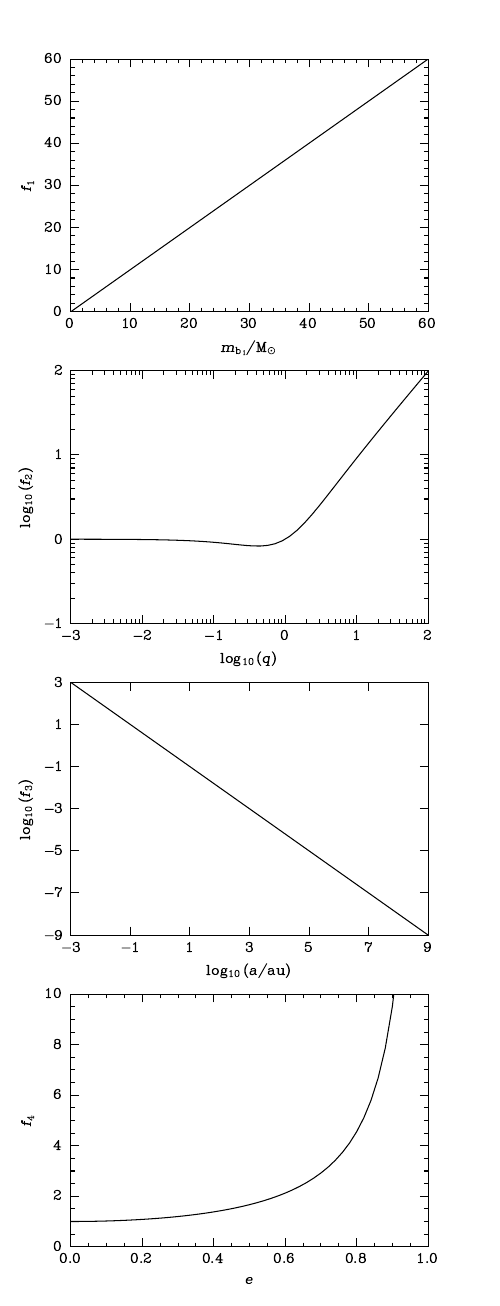}
\caption{\label{fig:org785cd72}The velocity dispersion of a population of identical binary stars is a function of primary mass, \(m_{\primary}\), mass ratio, \(q\), semimajor axis, \(a\), and eccentricity, \(e\), according to the formula \(\ms{V'_{b}} = f_{1}(m_{1})f_{2}(q)f_{3}(a)f_{4}(e)\) (Eq.~\ref{eq:inner_velocity_dispersion_factors}). Note that each factor of this formula may be arbitrarily large.}
\end{figure}

Since we have no closed-form expression for \(\ms{V^{\prime}_{\visual, z}}\), \(\ms{V^{\prime}_{\spectroscopic, z}}\), and \(\spectroscopicfraction\) in terms of the primary mass and orbital elements (as discussed in Sec.~\ref{org3341d43}) the effect of their distributions on the additional LOS velocity dispersion is not readily interpretable. However, we may find some insight into their effect by studying a single binary system with given primary mass, mass ratio, semimajor axis, and eccentricity. Consider the inner velocity dispersion of the primary star, which follows from Equation~\ref{eq:binary_component_velocity} and is
\begin{align}
\ms{{V'}_{\primary}} = \dfrac{\G{}q^{2}m_{\primary}}{a(1 + q)}\dfrac{1 + e^{2}}{1 - e^{2}},
\end{align}
while the inner velocity dispersion of the secondary star is
\begin{align}
\ms{V'_{\secondary}} = \dfrac{\ms{V'_{\primary}}}{q^{2}}.
\end{align}
The system's inner velocity dispersion is then
\begin{align}
\label{eq:inner_velocity_dispersion_factors}
\ms{{V'}_{\bin}}
=
\dfrac{\G{}}{2}
m_{\primary}
\dfrac{1 + q^{2}}{1 + q}
\dfrac{1}{a}
\dfrac{1 + e^{2}}{1 - e^{2}}.
\end{align}
This formula consists of four factors, each in one of the four variables \(m_{\primary}\), \(q\), \(a\), and \(e\). Let us call them \(f_{1}\), \(f_{2}\), \(f_{3}\), and \(f_{4}\). They are given by
\begin{gather}
\label{eq:factor_1}
f_{1}(m_{\primary})
= m_{\primary},\\
\label{eq:factor_2}
f_{2}(q)
= \dfrac{1 + q^{2}}{1 + q},\\
\label{eq:factor_3}
f_{3}(a)
= \dfrac{1}{a},
\end{gather}
and
\begin{align}
\label{eq:factor_4}
f_{4}(e) = \dfrac{1 + e^{2}}{1 - e^{2}}.
\end{align}
We plot these in Figure~\ref{fig:org785cd72}. 
Note that each of these may be arbitrarily large and that there is a minimum in \(f_{2}\) at \(q_{\min} = \sqrt{2} - 1\).

For our purposes Equation~\ref{eq:inner_velocity_dispersion_factors} tells us when the additional velocity dispersion is significant. In a \emph{binary system} this happens when
(1) the primary mass is large,
(2) the eccentricity is large,
(3) the semimajor axis is small, and
(4) the mass ratio is large.
In a \emph{population of binary systems} this happens when:
(1) there is an excess of high-mass stars,
(2) there is an excess of high-eccentricity systems,
(3) there is an excess of tight binaries, and
(4) there is an excess of inverted systems.

To understand the effect of our choice of distributions on the spectroscopic binary fraction, $\beta$, consider the time-averaged separation,
\begin{align}
\label{eq:average_separation}
\langle{}r\rangle
&= a\left(1 + \dfrac{e^{2}}{2}\right)
\end{align}
\cite[Eq.~1.65b][]{tremaine2023}.
This is an increasing function of semimajor axis and eccentricity.
In a \emph{binary system} it is small when
(1) the semimajor axis is small or
(2) the eccentricity is small.
In a \emph{population of binary systems} it is small when
(1) there is an excess of tight binaries and
(2) there is an excess of low-eccentricity systems.
In these circumstances we expect the spectroscopic binary fraction, \(\spectroscopicfraction\), to be large.

Equations \ref{eq:inner_velocity_dispersion_factors} and \ref{eq:average_separation} explain the fact that we see higher velocity dispersions when using a bottom-light initial mass function for our analysis of populations of ZAMS stars. 
In this scenario the average mass of the primary star is increased.
Since the distributions of mass ratio, semimajor axis, and eccentricity are conditional on primary mass so the average values of these quantities change too.
Specifically, 
the average value of the primary mass increases from $\qty{2.4}{\solarmass}$ to $\qty{7.2}{\solarmass}$, 
the average value of the mass ratio decreases from $0.49$ to $0.44$,
the average value of the semimajor axis increases from $\qty{440}{\astronomicalunit}$ to $\qty{490}{\astronomicalunit}$, and
the average value of the eccentricity increases from $0.54$ to $0.55$.
The effect of these changes overall is to increase both the inner velocity dispersion and average separation.
The dominant effect on the inner velocity dispersion is that of increased primary mass directly through $f_{1}$.
The dominant effect on the average separation is that of increased average semimajor axis.
This in turn reduces the number of spectroscopic systems, although the effect is marginal, in both cases the spectroscopic binary fraction being about $0.88$.
The two effects together conspire to increase the additional LOS velocity dispersion significantly.

Observations of clusters and galaxies are, of course, constrained by luminosity limits. Typically, surveys of the Milky Way's satellite galaxies are sensitive to stars at the tip of the RGB. We have assumed that the statistics of binary systems with RGB primaries of this kind are the same as those for systems with present-day solar-type primaries, as described by \cite{duquennoy1991}. Although this assumption is common in the literature it is not well justified. We would, ideally, know the statistics for those systems directly. If we wish to to use photometrically deeper observations of clusters and galaxies then we must in fact know the statistics for the whole RGB. We might assume that these are the same as the statistics of the ZAMS stars described by \cite{moe2017}. Ideally, however, we would know the statistics of binary systems drawn from all parts of the Hertzsrung--Russell diagram and choose those relevant to the analysis of data from any particular survey.    
These statistics can be determined by evolving a population of ZAMS binary stars using population synthesis \citep{izzard2018}. We plan to investigate this evolution in future work.

\subsection{The inner LOS velocity of a spectroscopic binary}
\label{sec:org0b0b4d1}
\label{org84b0e20}

The presence of binary stars always reduces the observed LOS velocity dispersion since the magnitude of the inner LOS velocity of a spectroscopic binary system, $v'_{\spectroscopic, z}$ (Eq.~\ref{eq:luminosity_weighted_los_velocity}), is always less than the magnitude of the LOS velocity of the primary star, ${v'}_{\primary}$.
The precise effect is dependent on the luminosity ratio, $\lambda := L_{\secondary}/L_{\primary}$ and the mass ratio, $q$. To see this let us rewrite Equation~\ref{eq:los_velocity} as
\begin{align}
\label{eq:standardized_primary_los_velocity}
{v'}_{\primary, z}
= \dfrac{q}{(1 + q)^{1/2}}A
\end{align}
where
\begin{align}
A := \left(\dfrac{\G{}m_{\primary}}{a(1 - e^{2})}\right)^{1/2}(\sin(i))(\cos(\nu + \omega) + e\cos(\omega)).
\end{align}
Then, by definition of the spectroscopic LOS velocity (Eq.~\ref{eq:luminosity_weighted_los_velocity}) and using Equation~\ref{eq:secondary_los_velocity} we have that
\begin{align}
\label{eq:spectroscopic_los_velocity}
v'_{\spectroscopic, z}
=
\dfrac{1 - \lambda/q}{1 + \lambda}
\dfrac{q}{(1 + q)^{1/2}}A.
\end{align}
Recall that, by definition, $\lambda \in (0, 1]$. There are two cases of interest: $q \in (0, 1]$ and $q \longrightarrow \infty$. 

\begin{figure}
\centering
\includegraphics[width=8.4cm]{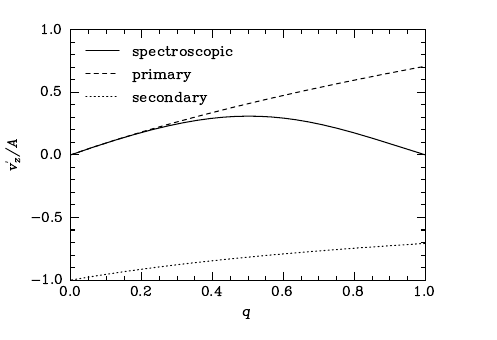}
\caption{\label{fig:orgc892b03}
The magnitude of the inner LOS velocity of a spectroscopic binary system is always less than or equal to the magnitude of its components' inner LOS velocities.
Here we show the inner LOS velocity of a spectropic binary system, $v'_{\spectroscopic, z}/A$ (Eq.~\ref{eq:spectroscopic_los_velocity}) assuming that stellar luminosity is a power law in mass with power-law index 3.5 (solid) alongside the LOS velocities of the primary star, ${v'}_{\primary, z}^{2}/A$ (dashed), and secondary star, ${v'}_{\secondary, z}^{2}/A$ (dotted).}
\end{figure}

For $q \in (0, 1]$ we indeed see that the magnitude of $v'_{\spectroscopic, z}$ is always less than the magnitude of ${v'}_{\primary}$. In fact for $q \ge \lambda$ it is always less then than both ${v'}_{\primary}$ and ${v'}_{\secondary}$. To illustrate this last point note that on the main sequence $q \in (0, 1]$ and that the luminosity may be approximated by a power law in mass with index $3.5$ so that $\lambda  = q^{3.5} \le q$. We plot $v'_{\spectroscopic, z}/A$ along with $v_{\primary, z}/A$ and $v_{\secondary, z}/A$ in Figure~\ref{fig:orgc892b03} adopting the convention that $v_{\primary, z}$ is nonnegative. The spectroscopic LOS velocity vanishes for $q = 1$ and as $q \longrightarrow 0$. It has a maximum at $q = 0.50$ (a value that depends on the chosen index of 3.5).

As $q \longrightarrow \infty$ so $v'_{\spectroscopic, z} \longrightarrow v'_{\primary, z}/(1 + \lambda)$ where $v'_{\primary, z} \longrightarrow \infty$. In other words, for very large mass ratios the spectroscopic LOS velocity is approximately equal to the LOS velocity of the primary star (which, in this case, is the less massive). This can be very large indeed. These inverted systems exist in populations of evolved binary stars, such as those with RGB primaries, although we exclude them from our analysis, as we have discussed.

\subsection{The validity of our assumptions}

Equations \ref{eq:pdf_los_velocity} and \ref{eq:los_stellar_velocity_dispersion}, where we began our analysis,  are completely general. They apply to clusters and galaxies of all kinds, including dwarf and disc galaxies. But Equations \ref{eq:mean_square_los_velocity} and \ref{eq:additional_mean_square_stellar_speed}, which form the basis for the rest of our analysis, rely on the assumptions that we discussed in Section 2.2, namely that (1) the outer velocity of any system is independent of its type (be it unary or binary) and (2) the inner velocities of the components of a binary system are independent of the outer velocity.
Both assumptions require the stellar population not to have undergone mass segregation. Since binary systems have higher masses than unary systems mass segregation would result in binary systems having a smaller spatial extent than unary systems. These binary systems will tend to have higher outer speeds, thus violating our first assumption. As we have just seen (Eq.~25), binary stars with high primary masses also have high inner speeds so that the most massive systems, which have the highest outer speeds, also have the highest inner speeds, thus violating our second assumption.

UFD galaxies have relaxation timescales that are longer than a Hubble time and therefore undergo no mass segregation. However, globular clusters have significantly shorter relaxation timescales meaning that the amount of mass segregation they undergo depends on their age \citep{baumgardt2022}. Some globular clusters have undergone mass segregation and some have not. We can therefore say that, in the case of UFD galaxies, our assumptions are not violated by mass segregation. But we cannot always say this about globular clusters.

Note that at no point in our analysis do we require that the potential be dominated by a single component (i.e.\ be dominated by the either the dark-matter halo or the stars). Our interest in UFD galaxies and globular clusters is motivated by our interest in near-field cosmology and not by the limitations of our methodology.

\subsection{A comparison with previously published results}
\label{sec:orgd7ebbdf}
\label{orgc772181}

The existing literature deals mainly with suites of small Monte Carlo samples of binary-rich star clusters, dSph, or UFD galaxies. The idea has been to determine the distribution of the LOS velocity dispersion as it would be computed using observational data for stellar populations of the appropriate sizes, i.e.\ for tens to hundreds of stars. In this approach a simulation of the whole cluster, dSph, or UFD galaxy (containing both unary and binary systems) is made.
Each system is assumed to have some outer LOS velocity drawn from a normal distribution with zero mean and variance equal to the intrinsic velocity dispersion of the system. The primary stars are invariably assumed to be RGB stars and therefore to have masses of \(\qty{0.8}{\solarmass}\). Morever, it is assumed that all binary systems are spectroscopic and that only the velocity of the primary star is observed. The inner LOS velocity of this primary star is drawn and added to the outer LOS velocity to give the LOS velocity in the galaxy's rest frame.

The most direct comparison is with the values of the LOS velocity dispersion for the spectroscopic binary population, \(\ms{V'_{\spectroscopic, z}}\), from which we can compute the additional LOS velocity dispersion for arbitrary binary fractions using Equation~\ref{eq:additional_mean_square_stellar_speed_unresolved}.
The literature tends not to contain explicit values for this but we can compute them using the published values of the observed LOS velocity dispersion and intrinsic LOS velocity dispersion using Equations~\ref{eq:mean_square_los_velocity} and~\ref{eq:additional_mean_square_stellar_speed_unresolved} since \(\ms{V'_{\spectroscopic, z}} = (\ms{V_{z}} - \ms{V_{\unary, z}})/\mathrm{B}\). In each case we will use the results for the largest binary fraction considered since the Monte Carlo simulations used to compute these contain the largest number of binary systems and are the most accurate.
We may then compare the literature values for $\ms{V'_{\spectroscopic, z}}$ with the value we have found, of $\qty{53.6}{\kilo\meter\squared\per\second\squared}$, by setting \(\alpha = 1\) and \(\beta = 1\) in Equation~\ref{eq:additional_mean_square_stellar_speed} for the case of systems with RGB primaries (see Fig.~\ref{fig:org7bdc51c}).

\subsubsection{Reported estimates of the LOS velocity dispersion}
\label{sec:orgfe2f4fd}

The first such studies were made by \cite{mateo1993}, in their analysis of Carina, and \cite{suntzeff1993} in their analysis of Sextans, using toy distributions of dynamical properties.\footnote{\cite{aaronson1987} had previously considered the case of multiepoch observations in their analysis of Draco and Ursa Major.}

\cite{mateo1993} considered an intrinsic LOS velocity dispersion of $\ms{V_{\unary, z}} = \qty{2.56(0)}{\kilo\meter\squared\per\second\squared}$ using binary fractions of $\alpha = 0.1$ and $0.2$.
In their analysis the distribution of eccentricities is uniform on the interval $e \in [0.5, 1)$, the distribution of periods is loguniform on intervals $P/\qty{}{\day} \in [0.5, P_{\max}/\qty{}{\day}]$ for $P_{\max} = 10, 100, 1~000$, or $\qty{10000}{\day}$, and the distribution of mass ratios is uniform on the interval $q \in [0.05, 1]$.
For $\alpha = 0.2$ they found the velocity dispersion to be between $\ms{V_{z}} = \qty{22.7(0)}{\kilo\meter\squared\per\second\squared}$ (when $P_{\max} = \qty{10000}{\day}$) and $\ms{V_{z}} = \qty{67.9}{\kilo\meter\squared\per\second\squared}$ (when $P_{\max} = \qty{10}{\day}$).
This is an additional velocity dispersion of $\deltams{V_{z}} = \qty{20.1}{\kilo\meter\squared\per\second\squared}$ and $\deltams{V_{z}} = \qty{65.3}{\kilo\meter\squared\per\second\squared}$.
Using Equation~\ref{eq:additional_mean_square_stellar_speed_resolved} we find that this is equivalent to a velocity dispersion for the binary population alone of $\ms{V'_{\spectroscopic, z}} = \qty{60.3}{\kilo\meter\squared\per\second\squared}$ and $\ms{V'_{\spectroscopic, z}} = \qty{196}{\kilo\meter\squared\per\second\squared}$.

\cite{suntzeff1993} considered an intrinsic LOS velocity dispersion of  $\ms{V_{\unary, z}} = \qty{4.41}{\kilo\meter\squared\per\second\squared}$ using binary fractions of $\alpha = 0.05, 0.1, 0.25$, and 0.5.
In their analysis the distribution of eccentricities is uniform on the interval $e \in [0, 1)$, the distribution of periods is unspecified but defined on the intervals $P/\qty{}{\year} \in [P_{\min}/\qty{}{\year}, 1000]$ for $P_{\min} = 0.5$ or $\qty{1000}{\year}$, and the distribution of mass ratios is uniform on the interval $q \in (0, 1]$. For $\alpha = 0.5$ they found that the velocity dispersion is between $\ms{V_{z}} = \qty{57.8}{\kilo\meter\squared\per\second\squared}$ (when $P_{\min} = \qty{1}{\year}$) and $\ms{V_{z}} = \qty{92.2}{\kilo\meter\squared\per\second\squared}$ (when $P_{\min} = \qty{0.5}{\year}$). This is equivalent to a velocity dispersion for the binary population alone of $\ms{V'_{\spectroscopic, z}} = \qty{80.1}{\kilo\meter\squared\per\second\squared}$ and $\ms{V'_{\spectroscopic, z}} = \qty{131}{\kilo\meter\squared\per\second\squared}$.

\cite{vogt1995} later considered an intrinsic LOS velocity dispersions of  $\ms{V_{\unary, z}}/\qty{}{\kilo\meter\squared\per\second\squared} \in [1, 100]$ for binary fractions of $\alpha = 0, 0.1, 0.2$, and 0.5.
In their analysis the distribution of eccentricities is uniform on the interval $e \in [0, 1)$, the distribution of periods is loguniform on the interval $P/\qty{}{\year} \in [0.5, 10~000]$ and the distribution of mass ratio unspecified.
For $\alpha = 0.5$ they found that the velocity dispersion is $\ms{V_{z}} = \qty{4.5}{\kilo\meter\squared\per\second\squared}$ when $\ms{V_{z}} = \qty{100}{\kilo\meter\squared\per\second\squared}$.
This is equivalent to a velocity dispersion for the binary population alone of $\ms{V'_{\spectroscopic, z}} = \qty{75}{\kilo\meter\squared\per\second\squared}$.

\subsubsection{Reported estimates of the additional line-of-sight velocity dispersion}
\label{sec:org267d91b}

More recently \cite{pianta2022} considered the expected velocity dispersion of binary-rich clusters using Monte Carlo simulations, arguing that binary contamination alone may reduce or eliminate the need for dark matter in explaining the kinematics of dSph and UFD galaxy candidates. They considered two star clusters, each representing the stellar population of a dSph or UFD galaxy in the absence of dark matter. In each case they modelled the cluster as a Plummer sphere, the dSph-like cluster having mass $M = 10^{7}\qty{}{\solarmass}$ and characteristic radius $b = \qty{3}{\kilo\parsec}$, with instrinsic velocity dispersion $\ms{V} = \qty{1.2}{\kilo\meter\squared\per\second\squared}$, and the UFD-like cluster having mass $M = \qty{5e4}{\solarmass}$ and characteristic radius $b = \qty{50}{\parsec}$, with an instrinsic velocity dispersion $\ms{V} = \qty{4.2}{\kilo\meter\squared\per\second\squared}$.

For each cluster they  considered a population of visual binary stars with stellar masses chosen using the random-pairing method assuming the canonical IMF of Kroupa with a mass interval bounded below by $\qty{0.1}{\solarmass}$ and above by $\qty{50}{\solarmass}$. Eccentricities were distributed thermally, and semimajor axes distributed loguniformly over the interval $a/\qty{}{\astronomicalunit} \in [a_{\min}/\qty{}{\astronomicalunit}, 100]$ for $a_{\min}/\qty{}{\astronomicalunit} \in [0.01, 1]$. They found that the expected velocity dispersion may be arbitrarily large given small enough $a_{\min}$. For example, using binary fraction $\alpha = 0.3$, if $a_{\min} = \qty{0.01}{\astronomicalunit}$ then both dSph-like and UFD-like clusters have an additional velocity dispersion of $\deltams{V} =  \qty{2.5e3}{\kilo\meter\squared\per\second\squared}$ (their Figs~2 and~5). Under the assumption of isotropy this is equivalent to an additional LOS velocity dispersion of $\delta{\ms{V_{z}}} = \qty{800}{\kilo\meter\squared\per\second\squared}$, significantly greater than our value of $\delta{\ms{V_{z}}} = \qty{180}{\kilo\meter\squared\per\second\squared}$. Such a large value would obviate the need for dark matter in UFD galaxies. However, the distributions of semimajor axes and eccentricities are loguniform and thermal only in the case of wide binary systems (where $a \gtrsim \qty{1e3}{\astronomicalunit}$). As a result there is a very large and unphysical excess of highly eccentric and tight binaries in the binary populations considered by \citeauthor{pianta2022}. Their results therefore reflect our finding (Sec.~\ref{sec:orgb6b717a}) that in such circumstances the additional velocity dispersion can be very large. 

\section{Conclusion}
\label{sec:org795f7f1}
\label{org7dd7a53}

We have described a formalism for modelling the LOS velocity distribution of galaxies and star clusters containing a population of binary stars and have applied it to UFD galaxies and globular clusters.
For a plausible binary fraction of 0.3 a fully resolved population of ZAMS binary stars contributes an additional LOS velocity dispersion of $\qty{180}{\kilo\meter\squared\per\second\squared}$.
However, a fully unresolved population contributes an additional LOS velocity dispersion of only $\qty{12.1}{\kilo\meter\squared\per\second\squared}$.
The effect of a mixed population of resolved and unresolved binary stars is subtle since the two subpopulations contain stars with different distributions of dynamical properties.
The unresolved subpopulation contains an excess of tight binaries with respect to the population of all binary stars, while the resolved subpopulation contains an excess of wide binaries.
For a given binary fraction the additional LOS velocity dispersion is therefore determined by the spatial resolution of our observations.
For spatial resolutions significantly smaller than the minimum allowed major axis, $2a_{\min}$, the population is \emph{effectively resolved}.
In this regime the additional velocity dispersion is large and approximately constant.
For spatial resolutions significantly greater than the minimum allowed major axis, $2a_{\min}$, the population is \emph{effectively unresolved}.
Here, the additional velocity dispersion due to binary stars is small and approximately constant.
All observations fall in this effectively unresolved regime.

For a binary fraction of 0.3 the resulting fractional mass increase for UFD galaxies is between 0.1 and 1. We might therefore expect to overestimate the masses of the coldest UFD galaxies by a factor of two.
The problem is more serious for globular cluster, where the fractional mass increase is between 0.1 and 100.
Moreover, the additional LOS velocity dispersion is sensitive to the distribution of primary-star masses, mass-ratios, semimajor axes, and eccentricities.
An excess of massive, tight, and higly eccentric systems can all increase the additional LOS velocity dispersion, as can an excess of inverted systems, in which the secondary star has a greater mass than the primary.
In particular, a bottom-light IMF of the kind described by \cite{geha2013} can increase the additional LOS velocity for both UFD galaxies and globular clusters by half a dex.

Our formalism allows for the possibility of fitting distribution-function models of UFD galaxies and star clusters directly to the data without the need for sample cleaning to remove binary pollution. Such a distribution-function model would include a description of the intrinsic dynamics and the dynamics of the binary population. However, the likelihood of its parameters would be multiply degenerate with the principal degeneracy being between host mass and binary fraction. A galaxy or cluster with given LOS velocity dispersion may be intrinsically heavy and binary poor or intrinsically light and binary rich. Moreover, small observational catalogues may not contain enough information to constrain these parameters. We leave this interesting problem to be addressed by future work.

\section{Acknowledgements}

We would like to thank the anonymous reviewer for the care and attention that they paid this paper. It is significantly better for their comments and suggestions.
AG and PD are supported by a UKRI Future Leaders Fellowship (grant reference MR/S032223/1). DDH is supported by a UKRI/UoS grant (reference H120341A).

\section{Data availability}

The computer code and data we have used to produce our results will be shared upon reasonable request to the corresponding author.

\bibliographystyle{mnras}
\bibliography{bibliography}

\end{document}